\documentclass[preprint,10pt,showpacs,prl,twocolumn]{revtex4}
\usepackage{amsmath}
\usepackage{amssymb}

\input{tcilatex}

\begin{document}

\preprint{}
\title[ ]{No new "Renormalized Magnetic Force Theorem'}
\author{V. P. Antropov}
\affiliation{Physics Dept., Ames Lab, Ames, IA, 50011}
\maketitle

Recently, in Ref.\cite{bruno} the author suggested a new \textquotedblleft
Renormalized Magnetic Force Theorem\textquotedblright\ for the calculation
of exchange interaction parameters and adiabatic spin-wave spectra of
ferromagnets. While this topic is of significant interest for the
computational magnetism community, the main results of this paper seem to be
not new, and their usage violates some general concepts of the magnetism
theory.

The following observations can be made:

1. Formally, the \textquotedblleft local force" theorem\ for spin rotations
from the Appendix of Ref. \cite{LIXT87} is formulated correctly and is
generally applicable. Although in the derivation of exchange parameters the
well known long-wave approximation (LWA)\cite{MORLWA,WANG} was used, but not
explicitly stated, the formulation of the theorem itself is \textit{correct}%
. The right numerical way to avoid LWA in the calculation of exchange
coupling within the local spin density approximation (LSDA) was first
suggested by Solovyev\cite{SOL}. It is evident that in Ref.\cite{bruno} the
author confused errors introduced by the LWA with errors in the force
theorem. His Eq.(11) is an expression for the exchange coupling beyond LWA
which does not constitute a new form of the "local force" theorem.

2. The central results of Ref.\cite{bruno} are given by Eqs.(15), (17) and
Eqs.(16), (18). As we have noted, these results are not new. Eqs.(15), (17)
follow the standard definition\cite{MORLWA} of the exchange coupling as the
inverse susceptibility, while Eqs.(16), (18) represent a transform of the
celebrated formula for the adiabatic spin wave spectrum in the rigid spin
approximation \cite{SWSAD}, clearly proving that the use of 'constraining
fields' concept does not introduce any new physics in the theory of
magnetism.

3. Eqs.(16), (18) are not generally applicable, and their limitations are
well documented\cite{ADIAB,gouteir}. In particular, the author claims that
the correct spin wave spectrum must be renormalized with a factor $(1-\omega
/\Delta )^{-1}$ (which results from the transformation of the exact
adiabatic formula to its LWA limit). However, all the calculations in this
paper were done in the adiabatic approximation which is also valid only in
the limit of $\omega \ll |\varepsilon ^{\uparrow }-\varepsilon ^{\downarrow
}|$. Since LWA and the adiabatic approximation are based on similar
smallness parameters, any improvement of LWA requires going beyond the
adiabatic approximation. Moreover, the random phase approximation (which was
used indirectly in Ref.\cite{bruno}) has analogous smallness criteria.
Therefore, both $\omega $ and $\widetilde{\omega }$ from Ref.\cite{bruno}
cannot be used at large $\mathbf{q}$ if $\omega /|\varepsilon ^{\uparrow
}-\varepsilon ^{\downarrow }|$ is not small. This statement is very general
and valid for any adiabatic theory.

The error is clearly seen in the calculation of $T_{c}$ in Ni. $T_{c}$ in
Ref.\cite{bruno} (630K) is in ideal agreement with experiment, but it
contradicts the LSDA itself. The LSDA total energy difference between the
ferromagnetic and non-magnetic states of Ni is about 520-550 K. These values
represent the upper limit for any reliable theory of $T_{c}$ based on LSDA.

4. The author claims that Eq.(7), which is the starting point for all
practical results in this paper, was derived in Ref.\cite{LIXT87}. This is
not true. Eq.(7) was not published previously; it contains intra-atomic
dispersion and, generally speaking, is incorrect (together with all formulas
where it is used). A somewhat similar expression (but in the rigid spin
approximation) was derived in Ref.\cite{WANG}, while in Ref.\cite{LIXT87} a
formula analogous to the result of Ref.\cite{WANG} was obtained within
multiple scattering theory.

The misunderstanding comes from the following statement: \textquotedblleft
\dots all results given here are straightforwardly extended to continuous
variables $\mathbf{u}(\mathbf{r})$, by omitting integration over atomic
cells $\Omega _{\mathbf{R}}$ everywhere, and replacing discrete summations
over $\mathbf{R}$ by continuous integrations over $\mathbf{r}$%
.\textquotedblright\ This claim is hardly justified. While the exact
formulation of the local force theorem contains the functional \textit{%
variation} of the total energy, in the rigid spin approximation (assuming no
intra-atomic dispersion) one has to use the corresponding \textit{derivative}%
: \cite{LANDAU7} 
\begin{equation}
\frac{\delta E}{\delta \mathbf{m}\left( \mathbf{R}_{i}+\mathbf{r}\right) }%
\approx \frac{\partial E}{\partial \mathbf{m}_{i}}+..  \label{c167}
\end{equation}%
with all gradient terms omitted (limit of large $\mathbf{q}$). Here $i$ is
the atomic site index. So far, this approximation was used in all papers on
exchange coupling calculations. Simply replacing the functional derivative
by the local derivative (as suggested by the author) can not be justified
beyond rigid spin approximation.

The correct formulas for the exchange coupling and spin wave spectrum in the
case when intra-atomic dispersion is important (full-potential treatment)
can be found, for instance, in Ref. \cite{gouteir}.

\end{document}